# Database Technology Evolution

Malcolm Crowe
Emeritus Professor, Computing Science
University of the West of Scotland
Paisley, United Kingdom
Malcolm.Crowe@uws.ac.uk

Fritz Laux
Emeritus Professor, Business Computing
Universität Reutlingen
Reutlingen, Germany
Fritz.Laux@reutlingen-university.de

***Abstract*– This paper reviews suggestions for changes to database technology coming from the work of many researchers, particularly those working with evolving big data. We discuss new approaches to remote data access and standards that better provide for durability and auditability in settings including business and scientific computing. We propose ways in which the language standards could evolve, with proof-of-concept implementations on Github.**

***Keywords*– *big live data; remote data; RDBMS; SQL; standards.***

## I. Introduction

The design of relational database management systems (RDBMS) has always focused on the management of structured and evolving data, such as customer accounts and scientific results, where shared access and long-term durability are important [1]. The Standard Query Language SQL, developed in the 1970s, rapidly became an international standard [2] with many features, and its evolution has been followed by most database products. Many researchers have been inspired to develop the theoretical underpinning for the implementation of these products, and this work continues today [3][4][5][6].

With all forms of evolution, some inherited aspects become awkward over time, for example, the early use of fixed-size fields and limited precision primitive types persists in database storage, limiting backwards compatibility of newer product versions and affecting durability and portability [7]. Some research projects including PyrrhoDB have chosen instead to use new globalized primitive types to avoid dependency on machine architecture and locale [8]. Avoiding such dependency facilitates data import and sharing, and the construction of data warehouses [9].

The development of data warehouses has led to a focus on metadata and semantics and has led many systems to use document-based NoSQL systems while other researchers have developed ways of including semantics in relational systems [10]. With these developments, it is natural to seek ways of adapting the relational DBMS paradigm to manage evolving data warehouse content (big live data) [11].

The tension between evolution and durability of *data* has always been a feature of relational database management systems (RDMS) and the associated technology. The use cases that inspired RDBMS development were business records such as customer accounts and inventories, and collaborative science, where support for shared access by many users with the responsibility for keeping data up to date needs to be balanced by the requirements for long-term storage, consistency, and audit. Over the years, such support has evolved, by the addition of powerful declarative and processing features in the evolving standard language SQL [2], and this evolution has come with a cost in compatibility between systems, since not all RDBMS implement the same version of the standard, and in durability, since RDBMS products also evolve, and not all RDBMS provide adequate backward compatibility to work with databases developed for a previous version. For these reasons, legacy data and systems are a continuing concern in all forms of business and scientific endeavor.

The starting point in this contribution is that the DBMS should generally support enterprise data integration where appropriate, and co-operative data sharing where this is useful. That is, the DBMS itself should support, but not require, ways of extending a data model through the enterprise, while providing mechanisms for supporting useful applications for the situations where the responsibility for data evolution is in another organization. In both cases the resulting structure will be a federation allowing some local management, with a hierarchy of delegation and responsibility, to avoid over-centralization on the one hand, or wasteful duplication on the other. This paper considers a number of improvements to DBMS technology designed to achieve this aim, while maintaining strong safeguards for preserving consistency for such complex systems where shared data evolves through supported activities in all parts of the system.

In the next section we consider an important set of use cases where people are interested in very targeted real-time data, gathered from many sources, where queries often lead to a unique entity on a single server. SQL remains a popular way of implementing database applications and even more general query systems, and ideally any changes should remain close to its original intent. In later sections of this paper, we examine some novel open-source approaches to such use cases in the PyrrhoDB project, which are based in widely used technologies and so have the potential to be useful in future big data developments. PyrrhoDB itself is a research project dating from before 2005 [12] rather than a product, but from its beginnings it has used globalized and machine-independent structures and the international standards and has always supported both evolution and backward compatibility.

In Section II we consider the state of the art, with an analysis of recent research papers that draw attention to changing requirements in database support for large and evolving data sets. This section also creates an agenda for the rest of the paper, to consider and suggest changes to relational data technology: *serialized* transactions and hierarchical privileges in Section III, proposals for the data type system and *metadata* in Section IV, *virtual* data warehousing (view-mediated remote access) in Section V, a suggestion to build implementations using *shareable* data structures in Section VI. Section VII looks at the implications for query processing, and Section VIII proposes a *versioned* API alongside the usual SQL data access methods and compares them with those of other database products. These sections include examples, and proof-of-concept implementations of these ideas are offered on Github.

## II. The use case of Big Live Data

Raw scientific and administrative data are often meaningless to the general public but is usually carried on the public web and usually has a significant real-time aspect.

Examples:

- The DNA signature of the latest Covid variants (whose data is progressively refined) [13],
- the latest data from sensors mapping a tsunami [14],
- the treatment history of a patient with a serious illness [15],
- the results from a particular fluid calculation that has taken a supercomputer three days to compute [16],
- the history of a piece of steel reinforcement in a tower block [17],
- the availability of intensive-care equipment for an emergency hospital admission [18],
- a particular sensor or actuator in the Internet of Things [19].

In some cases, there may be expectations coming from modeling (or AI) but a lot of important people in WHO, NASA, etc. want the scientists or investigators to get the right data. In some cases, the data (e.g., from sensors) is real-time, in others (e.g., the supercomputer example) the results may be a high-resolution image from numerical results that might not even be stored anywhere. Often such requests have life-and-death implications, and in order to guard against receiving approximate or out-of-date information, people resort to email or telephone.

In all such cases the data is conceptually part of a giant sparse database that no-one could possibly construct. Any individual observations would have lots of dependent metadata (provenance, device-specific details, confidence etc.). But often, the questions that the scientists want to ask are phrased in database terms, e.g., to examine the outcomes of patients with rare diseases and specific treatments, the quality of steel used in a component that needs to be replaced etc.

If SQL querying and secure remote update is also considered desirable, the above use cases point to some potentially desirable features. Excluding already-standard aspects such as authorization, universal time, international standards, auditing and linked data, and including features that not everyone would require, we can easily come up with the following wish list for SQL support:

- Search current data from a named collection of remote data sets
- Allow searching by metadata such as the resource description framework (RDF) or provenance where available
- Ensure transmitted data comes with timed provenance and ownership information
- Ensure remote updates (if permitted) are directly handled by the data owner, and fully recorded with user information of sender
- Avoid second-hand or out-of-date data by directly accessing the data's "transaction master"
- Specify service quality. e.g., to prioritize correctness over availability, report on out-of-date data or servers offline
- Minimize the amount of data that needs to be obtained or preloaded from remote servers
- Allow for transformation during retrieval, with inverses for updates if permitted
- Ensure changes are securely transacted, and durably recorded.

From the above discussion, in what follows we are motivated by the following general considerations:

- A focus on the need to support legacy data should motivate the separation of durable data from volatile data. The current state of any individual account or evolving record needs to be accessible from memory, but as in archiving, durable systems should prioritize and enable auditing of primary data such as particular inputs, changes, and deletions. In what follows, we reserve the concept of durable storage for this archive.
- On the other hand, access to and modification of the shared state of evolving data needs effective transaction control. The capturing of the desired durable archive then amounts to a log of such transactions, and the best way to prove the serializability of recorded transactions is that this log should itself record them atomically, in commit order, with all the steps of each commit kept together. Implementation of this log should be as append storage [20]. We note that some widely-supported DBMS features such as constraints, cascades and triggers complicate this requirement.
- Most DBMS are wary of the use of the Internet and prefer managing all network interaction using custom features. In our view this is now a mistake and ignores the opportunities for globalization that the evolving Internet standards offer. Greater opportunities for access should be balanced by better recording of data ownership, provenance, and responsibility, and these would help to address the concerns noted above for the ability in special cases to obtain results from (or even to update) sources rather than copies. We will demonstrate that such increased use of Internet standards has the potential to reduce wasteful data replication, especially for "live" data.

In considering the requirements for DBMS evolution, therefore, we consider the following aspects:

- The validation of transaction serialization, taking account of all side effects of transactions, so that transactions that violate constraints should not commit, nor if a resulting cascade or triggered action will conflict with other transactions. This requirement is mandated by the international SQL standard [2] but rarely implemented in commercial DBMS.
- We suggest a modified approach to DBMS design and security that places the data model and security model in the database rather than in applications. The SQL standard provides almost all of the support needed to achieve this: we take this forward by highlighting the definer's role for precompiled code and constraints, and through the creation of metadata features for the database itself. There are some consequential suggestions for enhancing SQL's extensive data type system.
- As in the US Department of Defense Orange Book standards for mandatory access control, we place the focus on user responsibility and security, while granting permissions to roles rather than users. Our proof-of-concept code includes the features required to implement the Orange Book levels B and C for users and database objects. Roles offer privileges on objects, and users are granted roles. We suggest however that the SQL standard should be modified so that a user can only use one role at a time. This is a practical suggestion since a user can be allowed to substitute for a sick colleague, but all actions are recorded in a way that identifies both the user and their declared role at the time.
- The SQL programming model is computationally complete: we recommend that the use of external code and procedures is disallowed, so that the DBMS can manage all of the validation and auditing required.
- In these circumstances, we support ways to allow better remote access to databases in SQL.

The remaining sections of this paper deal with practical proposals for all these aspects, making minimal changes to the SQL standard. Proof-of-concept code for these ideas already exists in PyrrhoDB on Github. Details are provided here in the following feature groupings: serialized transactions, DBMS accountability and data ownership, metadata, and view-mediated remote access.

## III. Serialized Transactions

From the above discussion, we implement a validation step for all transaction commits, to ensure that the requirement for fully serialized transactions is met. This renders obsolete the list of isolation levels (`READ_UNCOMMITTED`, `READ_COMMITTED`, `REPEATABLE_READ`, `SERIALIZABLE`) in the ISO standard, as there is only one possible isolation level, which could be called `SERIALIZED` [21],[22], reduces the number of available actions for integrity constraints by disallowing `NO ACTION` and limiting the extent to which constraints can be `DEFERRED`. The validation step guarantees fully isolated transactions. This means that changes made during a transaction are never visible to other users, but will prevent commit of conflicting transactions.

During a transaction, new records and database objects are temporarily given locations in memory, so that they are accessible and work as expected within the transaction thread. On commit, following the validation step, these objects are relocated in a cascade to the file positions where they will be recorded in the transaction log, and re-installed in the in-memory database. More details of this process are to be found in [23].

The granularity of the test for transaction conflict that is applied in this validation step is that (a) changes to the same database object (other than tables) will always conflict, (b) for tables, we report conflict if any columns read have been updated by another transaction, but if only specific rows have been read, we can limit the validation step to these rows. Validation for this level of granularity is practical even in situations of high concurrency [25]. The most recent implementation of this test (August 2022) uses two simple tree structures for columns and rows for any affected table, and also demonstrates correct behavior for cascades, constraints and triggers (files in [26] have been updated to show this).

For the best implementation of the optimistic concurrency control implied by the existence of the validation step in the commit algorithm, we advocate the use of shareable data structures. When discussing the sharing of modifiable data such as arrays, computer science textbooks often contrast the two approaches of copy on read and copy on write. From our point of view both are wasteful of time and resources, and the use of shareable data structures provides a different approach, which is well suited for the many tree-like structures found in database technology. A good way of motivating the concept is to consider the implementation of strings in programming languages.

In Unix, traditionally, strings (char *) are mutable: anyone with access to the string can modify individual characters in the string. In Java, C# and Python, strings are shareable: the only way to modify an individual character is to create a new string, so if a string is shared between two threads, any change to the string in one thread is not seen in the other thread unless it is explicitly given the new version.

Apart from strings, the most popular data structure in database technology is the B-tree, where each node apart from the root has at least n children and not more than 2n, where n>1, and information is placed in the leaves. In order to make database structures shareable, therefore, the key step is to use a shareable sort of B-tree. The model for this dates from 1982 [26], and the illustration reproduced in Figure 1 below shows that when a change to a tree is made to a leaf, we get a new root and the change requires $O(\log_n N)$ new nodes, where N is the number of leaves.

This means that the old and new version continue to share most of the nodes of the structure. With a little thought we can see that this is more storage-efficient than any of the approaches mentioned above (string implementation, copy on read, copy on write), but imposes a greater load on memory allocation and garbage collection. Crucially though, it is safe, and if we use this kind of structure for to implement all of the indexes and lists in the database many database operations such as starting a new transaction are made much simpler [27]. We return to these aspects in Section VII below.

The DBMS should specify and provide auditing support for a security model that allows local management. There is an opportunity for the SQL standard to encourage good practice in this area. PyrrhoDB has implemented the following practical steps for the local database:

*A. Maintenance of the full transaction log as the only artefact placed in non-volatile memory.*

There were good reasons for placing volatile information in non-volatile storage in 1972, but they are not valid now. It is understandable that where a database occupies large amounts of physical storage, a database administrator would regard the additional storage required for a transaction log as a luxury. PyrrhoDB's full transaction log is also serialized, so that it is evident that concurrent transactions have been correctly handled. Even in situations of high concurrency, the algorithms and solutions offered here have been shown to be practical [21].

When the only data written to disk is the inserted or updated record, or an indication that a record has been deleted, disk activity required for database traffic is drastically reduced, especially where the database has indexes that are stored on disk [12].

*B. Recording the user and role for each change to the database*

This is relatively easy to implement, though strongly resisted by database professionals and accountants, who dislike leaving their fingerprints all over the databases they administer or client account they prepare. However, it requires several departures from the SQL standard [2]: its features F771 and F321 allow the "current user" to be declared in the query language rather than being guaranteed by the operating system, and it does not demand that a user sets a single role. For forensic purposes, and to allow staff to substitute in different roles (due to illness etc) it is important to identify both user and declared role and is a simple matter if the transaction log is being maintained as suggested in Section III.A above.

In order to make the role and user information useful for forensic analysis, the grant of object ownership and role usage to roles should be deprecated, and the grant of anything other than these privileges to users should be deprecated.

As suggested in Section III.C, it should be possible to use the definer's role of an object to grant ownership to another user.

*C. Database objects should be modified only by their owner, and all execution should use definer's role*

From III.B, when objects are defined there is a current role: this is the definer's role, and it must be one of the roles that the user is permitted to use. This role and the owner's identity become properties of the object and can be modified by grant. The details of the new definition are checked both during parsing on every subsequent execution of the object.

The SQL standard specifies a context stack for procedure invocation, so it is again relatively easy to extend the use of such a stack for access to table columns, the sources of views, and the execution of constraints and triggers.

The execution engine then simply sets the current role for the called context to that of the definer of the table, view, procedure, constraint, or trigger, which it knows because of III.B. The invoker still needs appropriate permissions to initiate the process (by accessing or modifying the table or view or calling the procedure) and to access the columns of any table or row result.

The specifications in the standard make it very difficult to create a usable set of permissions for database operations, because users require usage permissions on every data type and column.

Two additional simplifications are recommended: the `REFERENCES` privilege in the standard then becomes redundant as it becomes the same thing as `SELECT`, and it simplifies the security model if all data types are usable by `PUBLIC` (though there may be restrictions on access to their fields if any). Using definer's role as described here, together with these changes, make the security model much easier to operate. New objects can be owned by the user that defines them (with their declared role as the definer's role) and the granting of privileges on an object does not need to consider data types or dependent definitions. Thus, it is much easier to maintain a usable set of privileges on even a large set of database objects.

With these provisions, Pyrrho's security model is simpler to administer and check for validity, but of course it makes execution somewhat slower: to check access permission on a single object requires a single access to the tree of properties of the object, which is typically of depth 3 (see below).

We believe this is an improvement on the arrangements used in Oracle [28] and PostgreSQL [29]. The cautionary words used about definer's role by these products are correct since they are installing native external procedures. Execution by the database server is safe because it can check all object permissions as they are accessed.

By using the role declaration model discussed above, all security settings for a relational database can and should be managed by the database itself, rather than in the database applications. The standard SQL model allows for hierarchical delegation of management of roles and permissions, separate from the authentication of users.

For example, consider the following simple database for a table-tennis club. It allows select access to the two tables shown, but changes to the database by ordinary members must be done with the help of the two procedures provided:

```
create table members (id int primary key, firstname char)
[create table played (id int primary key,
  winner int references members,
  loser int references members, agreed boolean)]
grant select on members to public
grant select on played to public
[create procedure claim(won int, beat int)
  insert into played(winner, loser)
    values(claim.won, claim. beat)]
[create procedure agree(p int)
  update played set agreed=true
   where winner=agree.p and loser in
    (select m.id from members m where user like
       '%'||firstname escape '^')]
create role admin
create role membergames
[grant execute on procedure claim(int, int) to role
```

```
  membergames]
grant execute on procedure agree(int) to role membergames
grant membergames to public
```

To use the given procedures, a member of the public who is allowed to login to the system should set their role to membergames.

## IV. The Type System and Metadata

A major difficulty in both enterprise data integration and data collaboration is the definition of a data model that supports application development in different parts of the enterprise. We consider it useful for databases to provide as much support for data semantics where possible, while retaining as much flexibility as possible for local development.

As a first step, we introduce the primitive Document type for JSON values and allow the braces '{' and '}' to delimit row values in SQL, the brackets '[' and ']' as string subscripts for Document values and a built-in Document-valued function HTTP whose parameters are the verb and url, with an optional third parameter being a Document for posted data.

Many DBMS have found the need to embellish their data access methods and database applications in various ways:

- Controlling XML and JSON output for queries, to identify whether table columns are output as attributes/fields or children/subdocuments of the table.
- For data visualization, e.g., charts
- Entity data models: Declaring classes in a database application corresponding to base tables in the database, with derived class references associated with foreign keys, lookup functions etc.

We consider it is good practice to include all such metadata in the database design, and it should be done on a per-role basis, to allow for suites of database applications for different business purposes.

In PyrrhoDB, we have come up with a list of useful metadata identifiers.

```
    Metadata  =  CAPTION | LEGEND  |  X | Y |
((HISTOGRAM | LINE | PIE | POINTS) ['(' id ',' id ')'])
    | ([URL | MIME | SQLAGENT | USER | PASSWORD] string)
| JSON | CSV | ETAG | MILLI
    | MONOTONIC | ((INVERTS|FORMATS) id)
    | ATTRIBUTE | ENTITY | ((SUFFIX|PREFIX) id) | iri .
```

This syntax is a Pyrrho extension, and metadata can be added to a database object (or dropped) by almost any DDL command. Most of the options affect query output for a role in Pyrrho's Web service. The above list provides a rough grouping of these keywords into four groups: (1) data visualization for specific tables and views, (2) provision for collaboration with remote data, (3) provision for adapter functions, and (4) support for local data models. `ATTRIBUTE` if present for a column indicates a preference for XML output for the containing table. `HISTOGRAM`, `LEGEND`, `LINE`, `POINTS`, `PIE` (for table, view or function metadata), `CAPTION`, `X` and `Y` (for column or sub-object metadata) specify JavaScript added to HTML output to draw the data visualizations specified. The syntax allows a string for a description. For `INVERTS` the id should be the name of the function being inverted, while for `FORMATS` the id is a type. `PREFIX` and `SUFFIX` define ids added to the client output string and in SQL triggers a default constructor for the type, as explained in the currency example at the end of this section.

Pyrrho helps with data visualizations defined using the keywords in group (1) above, using a simple URL-mapped HTTP service, as the following example shows:

With the database E created by

```
[create table sales (cust char(12) primary key,
custSales numeric(8,2))]
[insert into sales values ('Bosch' , 17000.00),('Boss'
,  13000.00), ('Daimler',20000.00)]
[insert into sales values
('Siemens',9000.00),('Porsche', 5000.00), ('VW',
8000.00), ('Migros' , 4000.00)]
create role E
grant E to "usermachine/username"
```

The data visualization output uses HTML returned to the client application or for immediate display. Here, if the browser is asked for

```
http://localhost:8180/E/E/SALES/?PIE(CUST,CUSTSALES)LEGEND
```

The browser will display the following output from the PyrrhoDB server:

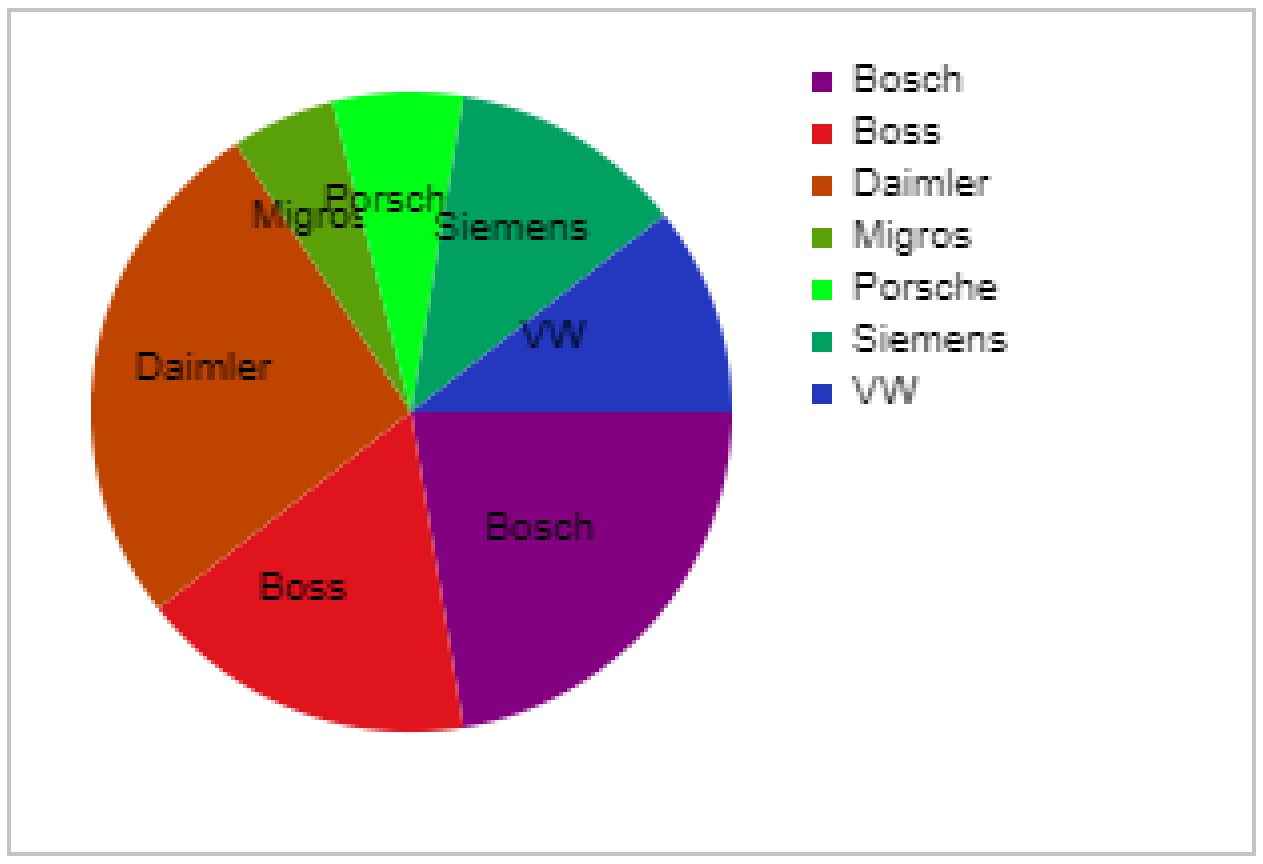

We return to this example below.

User-defined types can nominate a primitive type in the `UNDER` clause, and this can be useful for distinguishing data that has been imported or used in different suborganisations. The SQL standard already provides the `OF` predicate for selecting a value of a type, a `TREAT` function for specifying the subtype for a scalar value, and a "create table of type" mechanism for specifying row types. Pyrrho adds the ability to specify a subtype for `VALUES`.

As an example of the resulting syntax, if we defined:

```
[create type currency as(amt numeric,unit char)
  method exchange(tounit char) returns currency,
  method tonumeric() returns numeric]
```

The exchange method here would be implemented for the database using the above-mentioned HTTP function. There are many currency converters available on the Internet, for example

```
[create method exchange(tounit char) returns currency
for currency
    begin
      if unit=tounit return this;
      declare rates document;
      declare roe numeric;
      set rates=http('post',
'http://www.floatrates.com/daily/'||unit||'.json');
      set roe=rates[lower(tounit)]['rate'];
      return currency(amt*roe,tounit)
    end]
```

Then we could have

```
[create type dollars under currency check(unit='USD')
   constructor method dollars(x numeric),
   constructor method (x currency) prefix "$"]
[create constructor method dollars(x numeric)
   begin set amt = x; set unit = 'USD' end]
[create constructor method dollars(x currency)
   begin set amt=x.exchange('USD').amt;
   set unit='USD' end]
```

If we have similar declarations for euros, we could write things as simple as

```
select euros("$" 10)
create table money (cur currency)
insert into money values ("$" 34), ("€" 212.7)
select * from money where cur is of(dollars)
```

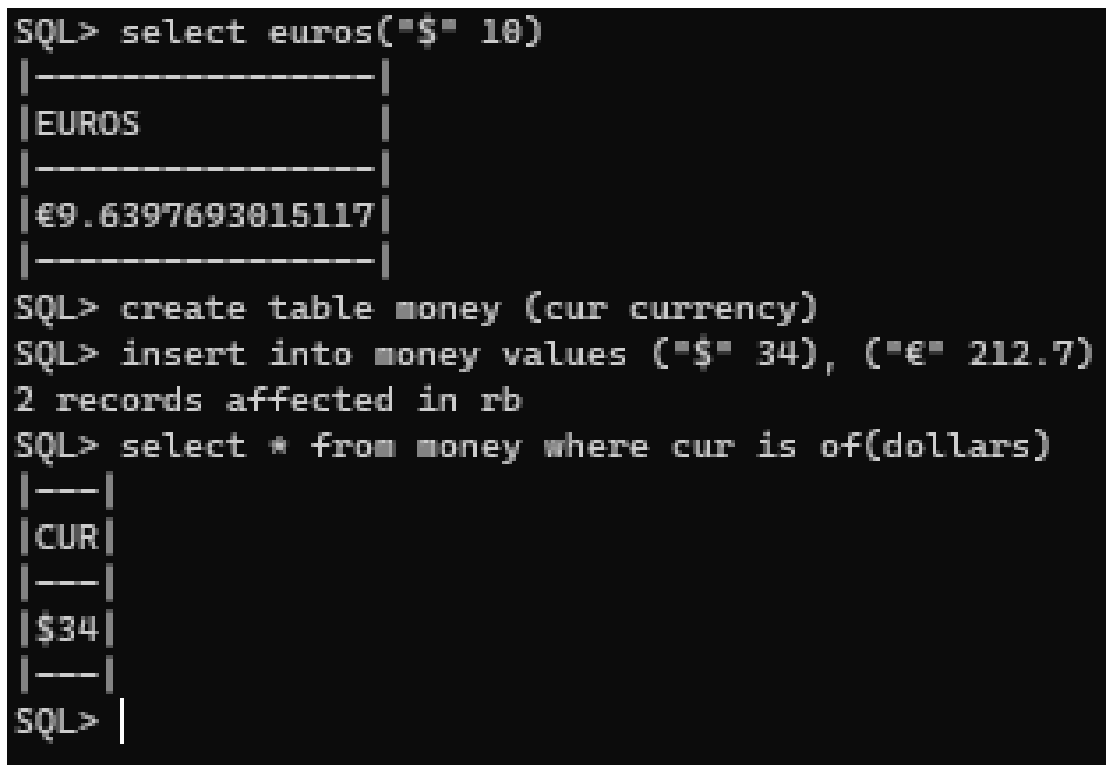

We give an example using the data model metadata directive `ENTITY` in Section VIII below.

## V. VIEW-MEDIATED REMOTE ACCESS

Data warehousing involves creating central data repositories (using extract-transform-load technologies) to enable analytic processing of a combined data set. There are several situations where this is undesirable, for example where the resulting data protection responsibility at the central repository is excessive, where the data is volatile and it becomes expensive to maintain all of the centrally-held data in real time, or where it is better to leave the data at its sources where the responsibility lies [10]. With database technology, a View (if defined but not materialised) allows access to data defined in other places. The virtual data warehouse concept exploits this notion, and endeavours to avoid the central accumulation of data. Pyrrho uses HTTP to collect data from the remote DBMS using a simple REST interface [22], and so the resulting technology here is called RESTView.

Thus, with RESTView, a Pyrrho database allows definition of views where the data is held on remote DBMS(s), and is accessible via SQL statements sent over HTTP with Json responses. Pyrrho itself provides such an HTTP service and the distribution includes suitable interface servers (RestIf) to provide such a service for remote MySQL and SqlServer DBMS. The implementation allows for authentication as an ordinary client of the remote DBMS, whose administrator can grant access to a suitably defined view.

The HTTP access provides the user/password combinations set up for this purpose within MySQL by the owners of contributor databases. In the use cases considered here, where a query Q references a RESTView V, we assume that (a) materializing V by Extract-transform-load is undesirable for some legal reason or because of the high data volumes required, and (b) we know nothing of the internal details of contributor databases. A single remote select statement defines each RESTView: the agreement with a contributor does not provide any complex protocols, so that for any given Q, we want at most one query to any contributor, compatible with the permissions granted to us by the contributor, namely grant select on the RESTView columns.

Crucially, though, for any given Q, we want to minimize the volume D of data transferred. We can consider how much data Q needs to compute its results, and we rewrite the query to keep D as low as possible. Obviously, many such queries (such as the obvious select * from V) would need all of the data. At the other extreme, if Q only refers to local data (no RESTViews) D is always zero, so that all of this analysis is specific to the RESTView technology.

During query processing views are replaced by their definitions, so that the overall query becomes a selection from the tables they reference. The process deals with the situation that a table can be referenced in more than one place by adding unique identifiers for each table reference.

Filters are applied at the lowest level of the query (e.g., directly on a remote table), and traversal of a remote table creates a roundtrip of the REST service to the given URL. The JSON representation of the result returned is slightly enhanced to add the registers used to compute any remote aggregations [23].

The syntax is

ViewDefinition = [ViewSpec] AS
    (QueryExpression | GET [USING Table_id]) {Metadata}.

The alternative shown by the vertical bar corresponds to whether the view has one single contributor or multiple remote databases. The QueryExpression option here is the normal syntax for defining a view. The REST options both contain the GET keyword. The simplest kind of RESTView is defined as GET from a url defined in the Metadata. The types of the columns need to be specified in a slightly extended ViewSpec syntax. If there are multiple remote databases, the GET USING table_id option is available. The rows of this table describe the remote contributions: the last column supplies the metadata for the contributor including a url, and data in the other columns (if any) is simply copied into the view. For example:

```
SQL> create view VV of (E int,F char) as get 'http://localhost:8180/DB/DB/t'
SQL> create view WW of (E int, D char, K int, F char) as get using VU
SQL> select * from ww where e=5
|-|-|-|----|
|E|D|K|F   |
|-|-|-|----|
|5|C|1|Five|
|-|-|-|----|
SQL> table vu
|-|-|-----------------------------|
|D|K|U                            |
|-|-|-----------------------------|
|B|4|http://localhost:8180/DB/DB/t|
|C|1|http://localhost:8180/DC/DC/u|
|-|-|-----------------------------|
SQL> |
```

Depending on how the remote contributions are defined, RESTViews may be updatable, and may support insert and delete operations.

The implementation of these ideas was demonstrated in [23].

With these arrangements it is important to consider transaction requirements for multiple-host scenarios. The fundamental difficulty is the so-called two-army problem, according to which all data needs a single transaction master. Every transaction is initiated at one database (call its server's host local), and then accesses remote data via a view definition of the type described above. The transaction can commit changes on the local server and at most one remote server update, assuming the transaction provides suitable credentials for that database. The commit takes place according to the following mechanism (a) the local database is locked, (b) the local changes are validated, (c) HTTP 1.1 is used to perform the single remote update (using the RFC7232 mechanisms), (d) then the local commit can complete and unlock. With just one remote update this mechanism is safe and can be rolled back on any exception.

It is possible to imagine interworking between heterogeneous DBMS using these techniques, so that it is important to maintain the use of standard industry approaches for REST services. Many systems have implemented a URL and XML/JSON to database mapping, and the ETag mechanism from RFC7232 [24] can be leveraged to provide transactional features [20]. Currently in Pyrrho there are several options for this, determined by the metadata flags `URL` and `ETAG` listed above.

Consider again the sales database E from Section IV, which over time gains a great many sales records. Suppose E offers to role rs_V a view into the data that includes a computation of the current runningSalesShare as a number between 0 and 1:

```
[create view sales_V(cust, custSales, runningSalesShare)
 as select cust, custSales,
  (select sum(custSales) from sales where custSales >=
u.custSales) /
   (select sum(custSales) from sales)
from sales as u]
create role rs_V
grant rs_V to "user\machine"
```

Then this view can be accessed from the named machine using dashboard-style queries that categorize the customers A, B or C depending on the current runningSalesShare without having to be told all of the individual sales.

```
[select case when runningSalesShare <= 0.5 then 'A'
  when runningSalesShare > 0.5  and
   runningSalesShare <= 0.85 then 'B'
  when runningSalesShare > 0.85 then 'C'
  else null
  end as Category,
 cust, custSales,
 cast(cast(custSales  /  (select  sum(custSales)  from
sales_V) * 100
   as decimal(6, 2))
  as char(6)) || ' %' as share
from sales_V
order by custSales desc]
```

The output, and a pie chart derived from it, are shown in Figure 2.

## VI. Implementation using shareable data structures

This section provides some details of the implementation for the above features, following the philosophy outlined above using globalized and architecture-independent data formats. PyrrhoDB uses 64-bit uids for all database objects, log entries, and table rows. It uses a representation for variable length primitives Integer (up to 2040 bits), Real (Integer mantissa, int scale) and Char (Unicode strings up to $2^{60}$ bytes). The naming of database objects (except data types) is on a per-role basis.

As mentioned above, the database is represented on disk by a transaction log, consisting of a sequence of "physicals": there are roughly 70 physical formats: one of these is for transaction details, another for a table identifier, another for column details etc. The details are in the Pyrrho manual in the Github distribution. The transaction log uses append storage.

On first access by the server a database's entire transaction log is read and the live database objects constructed in memory.

Following the success of StrongDBMS [21] in performing serializable transactions in a high-concurrency demonstration, PyrrhoDB has been re-implemented to use shareable data structures throughout. A shareable data structure cannot be updated or modified, so any change involves creation of a new instance. Examples of shareable data structures are primitive data types such as integer or float, the string type in C#, Java or Python, and classes whose fields are all readonly shareable data structures. With the help of some simple shareable building blocks (BList<V> and BTree<K,V>) it is straightforward to build up shareable data structures representing tables, indexes and even databases.

A Domain class specifies a base type and many other properties. If it has columns (e.g., a user-defined type or base table) the Domain will also specify a list of column uids, and a tree giving the Domain of each field. The primitive types have system-allocated (negative) uids, and any other Domain is created as physical objects in the database that defines it.

Table rows are composed of TypedValues: a TypedValue is defined by a Domain and a shareable data structure.

The BTree<K,V> implementation was described above: it is an unbalanced B-Tree that gives worst-case O(logN) performance for inserting, changing or deleting a node. Any of these changes creates a new root node and new internal nodes to the new leaf node, making at most logN new nodes, while the rest of the nodes are shared between the old and new version of the tree. This is therefore surprisingly efficient.

BList<V> is not so clever. It is implemented as a BTree, but it renumbers its nodes 0, 1, 2, … resulting in a worst-case performance of O(N).

BList and BTree are shareable data structures provided their contents (all K and V objects) are shareable. Instead of the enumerators found in Java and C#, traversal of BTrees and BLists uses "bookmarks" that are also shareable data structures: two-way traversal is possible, and traversal continues to traverse from the root it was given and so is unaffected by changes to the tree it is traversing.

All classes that make up database objects are shareable. For example, Rowsets are basically a BTree of rows that are TypedValues, traversed by Cursors, which are a subclass of the bookmark class mentioned above.

A new server thread is started for each connection to a database. Protocol requests typically create a Transaction to query or modify the database or read the next group of data from the result of a query, which is confined to the connection thread The creation of a transaction is a simple matter: each transaction starts with a copy of the root node of the database (a snapshot). On rollback or disconnect, the transaction can simply be forgotten, as no other thread has seen it.

## VII. Query processing and Compiled objects

During parsing, uids are allocated for the resulting expressions, and for anything that may be committed as a new database object. Uids in the range above (currently) $4\times2^{60}$ are allocated as required: there are several ranges for these depending (for example) on whether their lifetime is the current session, the current transaction, or the current lexical input. SQL expressions all have Domains discovered during parsing, and RowSets all have Domains that specify their columns, so that ad-hoc Domains are constructed as required during query processing. Since a query may reference a table source more than once (via a TableRowSet), the column uids for TableRowSets need to be specific to such a reference and are allocated in the heap range (above $7\times2^{60}$): this process is called instancing in the implementation. Views may also reference more than one source, so the instancing process also applies to them. A similar requirement exists for table-valued functions.

In this section we also consider how the concept of local data management can be realized. As mentioned in Section III.C, the server should remain in control of execution of stored procedures, triggers, and constraints, so that such features should be written in SQL. Since the definer of a compiled object generally has different privileges from the user making a query or update, it is important to ensure that executable code is compiled in advance. For reasons of forward and backward compatibility the database file contains only the SQL source code for stored procedures, constraints, triggers, etc. The compiled components are constructed when the database is loaded in the server (after a cold start).

As mentioned above, many database objects correspond to permanent physical records in the transaction log, and so their defining position is fixed and they can be shared with all transactions for this database. Objects constructed by the server during compilation (also in fact shareable) do not have physical file positions, so instead receive uids in a dedicated range (currently $6\times2^{60}$ .. $7\times2^{60}$-1), and form a collection stored with the in-memory version of the compiled object. Most compiled objects contain executable code, but this mechanism is also used for the Domain of a base table or view. The actual uids allocated to these compiled objects will depend on the order of the physical objects in the log, and will depend on the current version of the server. During instancing, column uids will be allocated in a cascade, since many compiled objects will contain references to the columns being instanced.

Cascades are also used in the process of RowSet review, in which the RowSet pipeline is simplified wherever possible based on the existence of indexes and filters that were not available at compilation time.

## VIII. The Versioned library and data models

The above discussion described how the data model for a database could be represented in the database implementation. The real benefit of placing the data model in the database is to make it available to the application programmer, so that all applications targeting a database can agree on the structure and semantics of its data. At present, Pyrrho provides such support for applications written in C#, Java, and Python, in addition to a thread-safe version of the Command/ExecuteReader/Read programming interface familiar from ADO.NET and JDBC.

The current implementation was inspired by Microsoft's Entity Framework [30] and Java Persistence Architecture [31] but differs from these in the crucial proviso that application programmers should start with class definitions generated by (and at runtime checked by) the server, rather than writing their own version of the model in the form of annotations or code attributes.

The following example illustrates the type of support available. Suppose a database ABC contains a role "Sales" that defines the following tables:

```
[create table "Customer"(id int primary key, "NAME"
       char unique)]
[create table "Order"(id int primary key, cust int
       references "Customer", "OrderDate" date, "Total"
       numeric(6,2))]
```

Then the system table "Role$ClassValue" will provide code fragments similar to the following:

```
using System;
using Pyrrho;

/// <summary>
/// Class Customer from Database ABC, Role Sales
// PrimaryKey(ID)
// Unique(NAME)
```

```
/// </summary>
[Table(23,122)]
public class Customer : Versioned {
[Field(PyrrhoDbType.Integer)]
[AutoKey]
  public Int64? ID;
[Field(PyrrhoDbType.String)]
  public String? NAME;
  public Order[] orders =>
     conn.FindWith<Order>(("CUST",ID));
}
/// <summary>
/// Class Order from Database ABC, Role Sales
// PrimaryKey(ID)
// ForeignKey, RestrictUpdate, CascadeDelete(CUST)
/// </summary>
[Table(175,362)]
public class Order : Versioned {
[Field(PyrrhoDbType.Integer)]
[AutoKey]
  public Int64? ID;
[Field(PyrrhoDbType.Integer)]
  public Int64? CUST;
[Field(PyrrhoDbType.Date)]
  public Date? OrderDate;
[Field(PyrrhoDbType.Decimal,"Domain   NUMERIC   Prec=6
Scale=2")]
  public Decimal? Total;
  public Customer customer =>
     conn.FindOne<Customer>((“ID”,CUST));
}
```

The numbers 23 and 175 are references to the defining positions of these objects in the database, and the other numbers are schema keys, which will be checked by the server when the application runs to ensure that the table definition has not changed. We can see that the columns defined for the table are publicly accessible in these classes (while the server will check the user and role on access), and Pyrrho's data types of these columns are provided as attributes.

Importantly, the foreign key relationship between the tables has resulted in two additional "navigation" fields in the classes above, providing quick access to the customer for an order, and the orders for a customer. The primary and unique key declarations also allow quick access.

A simple program to use the above class definitions could begin

```
static void Main
{
 conn = new PyrrhoConnect("Files=Demo;Role=Sales");
 conn.Open();
 try
 {
// Get a list of all orders showing the customer name
   var aa = conn.FindAll<Order>();
   foreach (var a in aa)
    Console.WriteLine(a.ID + ": " + a.customer.NAME);
   if (aa.Length == 0)
   {
    Console.WriteLine("The Order table is empty");
    goto skip;
   }
 // change the customer name of the first
 // (update to a navigation property)
   var j = aa[0].customer;
   j.NAME = "Johnny";
   j.Put();
 // add a new customer (autokey is used here)
   var g = new Customer() { NAME = "Greta" };
   conn.Post(g);
 // place a new order for Mary
 // (secondary index, single quotes optional here!)
   var m = conn.FindOne<Customer>(("NAME","Mary"));
   var o = new Order()
    { CUST = (long)m.ID,
      OrderDate = new Date(DateTime.Now) };
    conn.Post(o);
```

The Versioned base class above uses ETags, allowing the library to associate object references in the code to rows in the database, and this enables the shorthand notation in the above sample program, in addition to providing automatic transaction validation when committing an explicit transaction is started, using an API similar to ADO.NET and JDBC. For further details see the Pyrrho manual [32].

## IX. Conclusions

This paper has reviewed a number of desirable changes to the relational database model that have been signaled in recent literature and outlined implementations of these improvements that can be found in the ShareableDataStructures project on Github [32]. The implementation of PyrrhoDB v7 is currently at the alpha stage and feedback on these ideas is welcomed. The authors are grateful for the many expressions of support and encouragement we have received during this project.

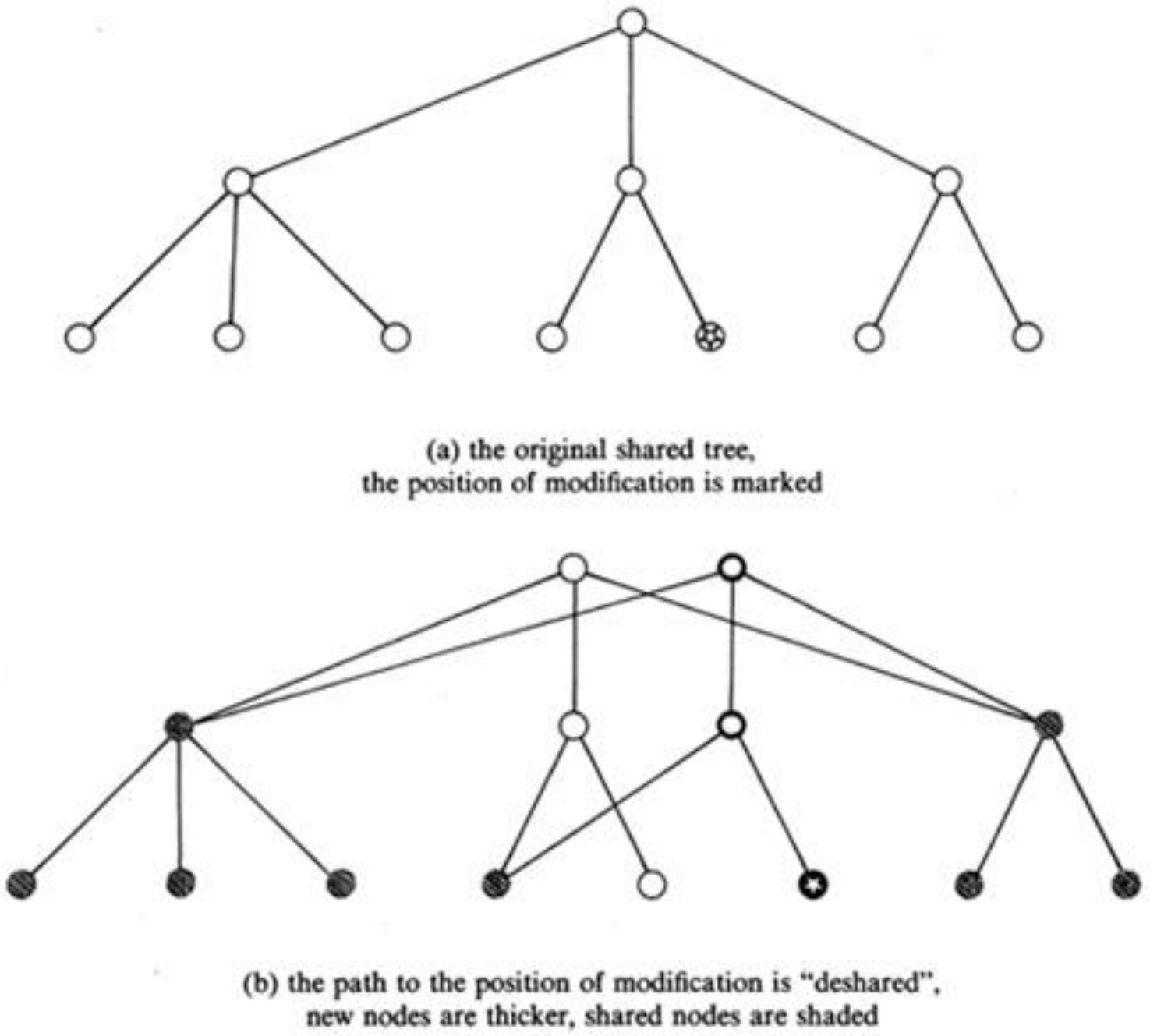

Figure 1. Operation of B-Trees [26]

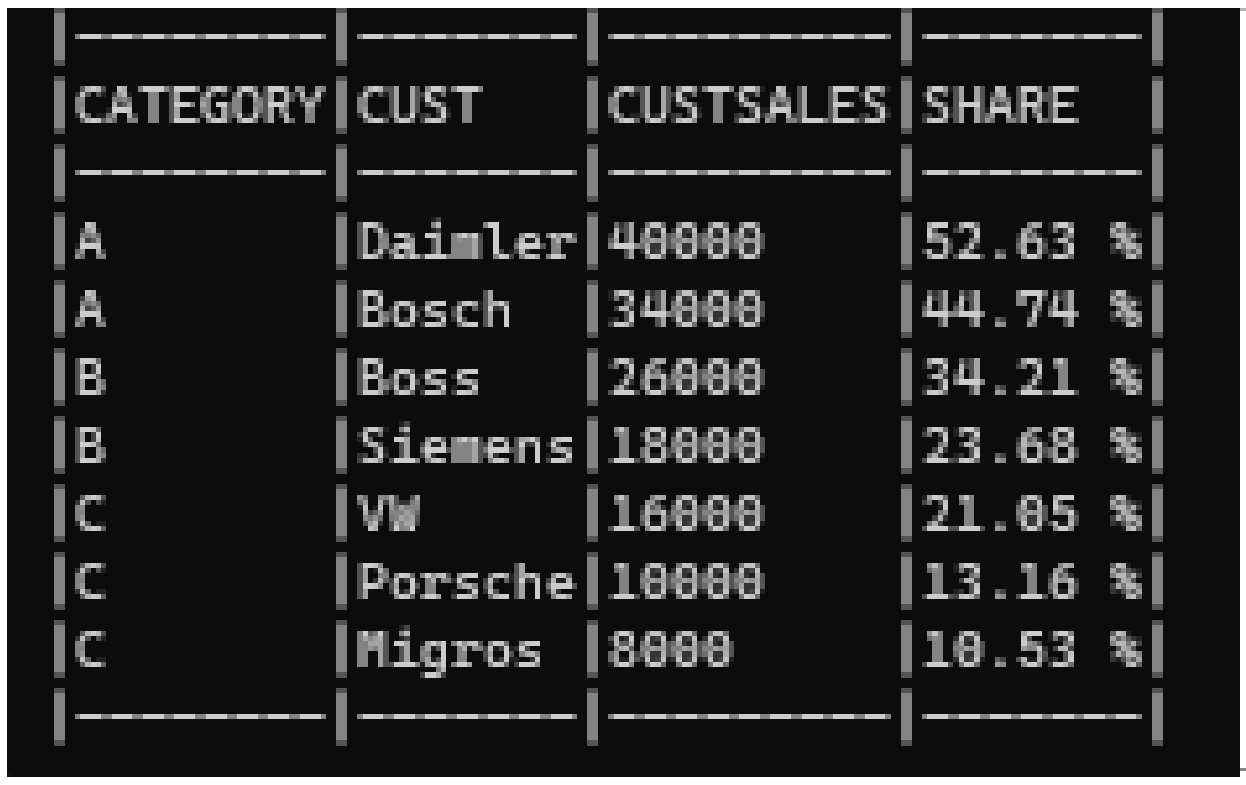

| CATEGORY | CUST | CUSTSALES | SHARE |
|---|---|---|---|
| A | Daimler | 40000 | 52.63 % |
| A | Bosch | 34000 | 44.74 % |
| B | Boss | 26000 | 34.21 % |
| B | Siemens | 18000 | 23.68 % |
| C | VW | 16000 | 21.05 % |
| C | Porsche | 10000 | 13.16 % |
| C | Migros | 8000 | 10.53 % |

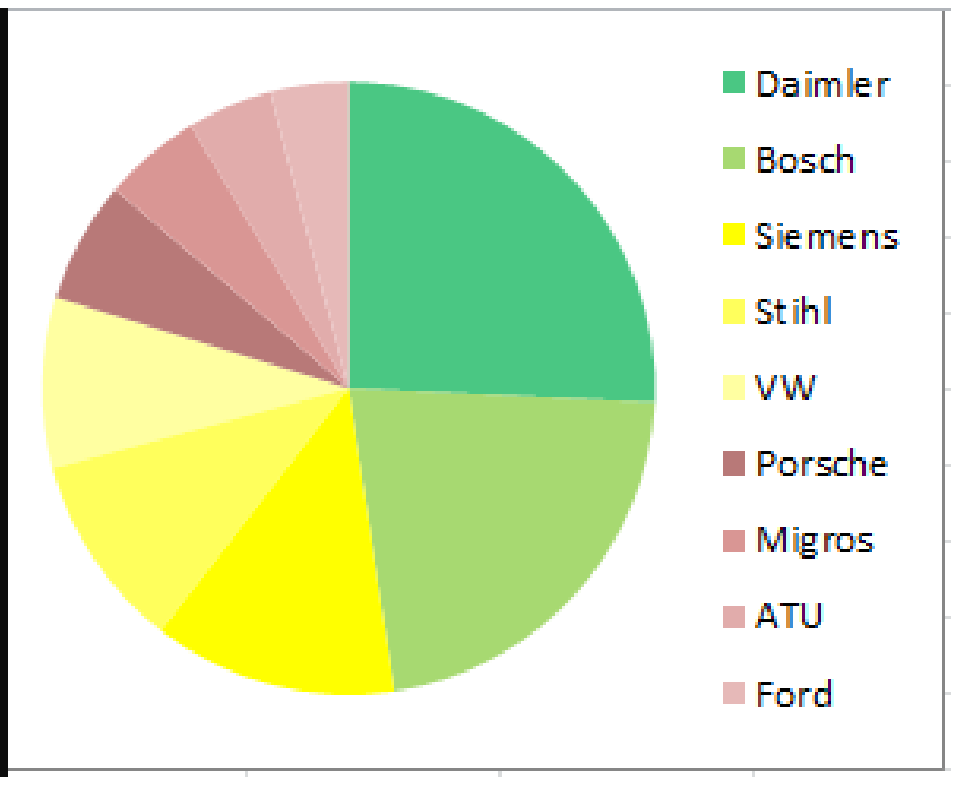

Figure 2: ABC analysis from Section VI example (as output from PyrrhoDB client and server)